\documentclass{article}
\usepackage{epsfig,graphicx}   
\usepackage{float}    
\usepackage{amsmath}
\usepackage{authblk}
\usepackage{xcolor}
\usepackage{enumitem}

\title{An Abelian Higgs model of pulsed field magnetization in superconductors}

\author[1]{J. G. Caputo \thanks{caputo@insa-rouen.fr}}
\author[2]{I. Danaila \thanks{ionut.danaila@univ-rouen.fr}}
\author[2]{C. Tain \thanks{cyril.tain@univ-rouen.fr}}

\affil[1]{Laboratoire de Math\'ematiques, INSA de Rouen Normandie\\ 76801 Saint-Etienne du Rouvray, France.}
\affil[2]{Laboratoire Raphael Salem, Universit\'e de Rouen Normandie, CNRS UMR 6085, 76800 Saint-Etienne du Rouvray, France. .}

\date{ }

\begin{document}
\maketitle

\begin{abstract}

Pulsed field magnetization leads to trapped magnetic field persistent for long times.We present a one-dimensional model of the interaction between an electromagnetic wave and a superconducting slab based on the Maxwell-Ginzburg-Landau
(Abelian Higgs) theory. We first derive the model starting from a Lagrangian coupling the
electromagnetic field with the Ginzburg-Landau potential for the
superconductor. Then we explore numerically its capabilities by applying a Gaussian vector potential pulse and monitoring usual quantities such as the modulus and the phase of the order parameter. We also introduce defects in the computational domain. We show that the presence of defects enhances the remanent vector potential and diminishes the modulus of the order parameter, in agreement with existing experiments.

\end{abstract}

\section{Introduction}

\begin{figure}[h]
\begin{minipage}{14pc}
\centering
\includegraphics[height=4.9 cm,width=6.1 cm]{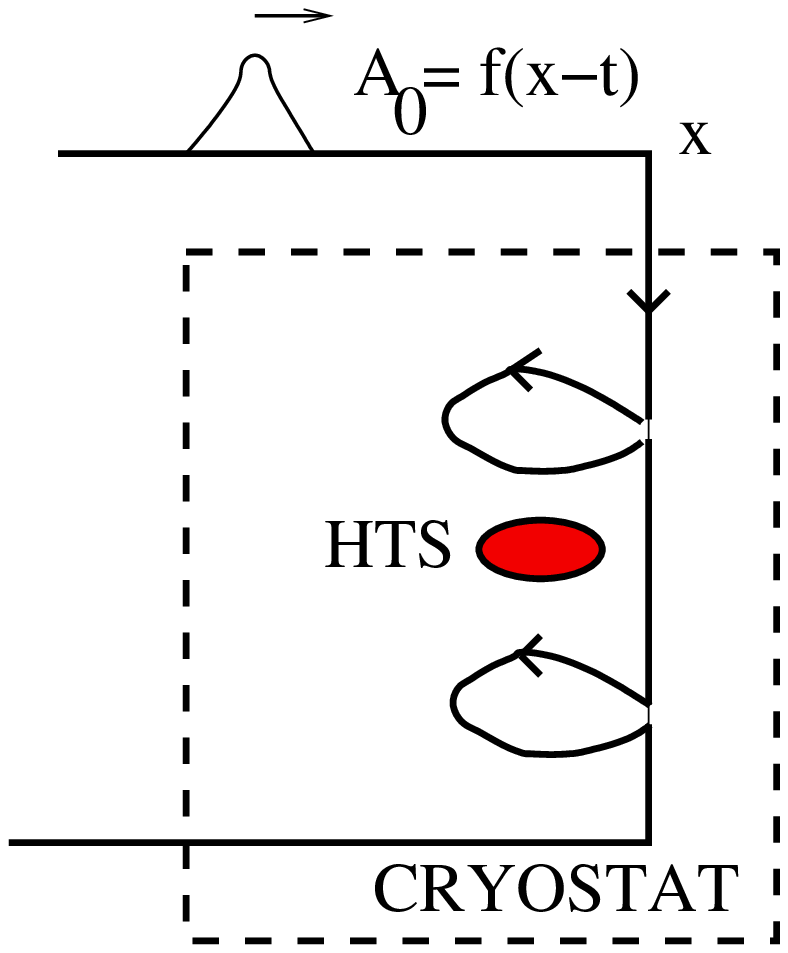}
\caption{\label{fpm1}Pulse field magnetization experimental set up \cite{weinstein}.}
\end{minipage}\hspace{2pc}%
\begin{minipage}{14pc}
\centering
\includegraphics[height=4.9 cm,width=6.1 cm]{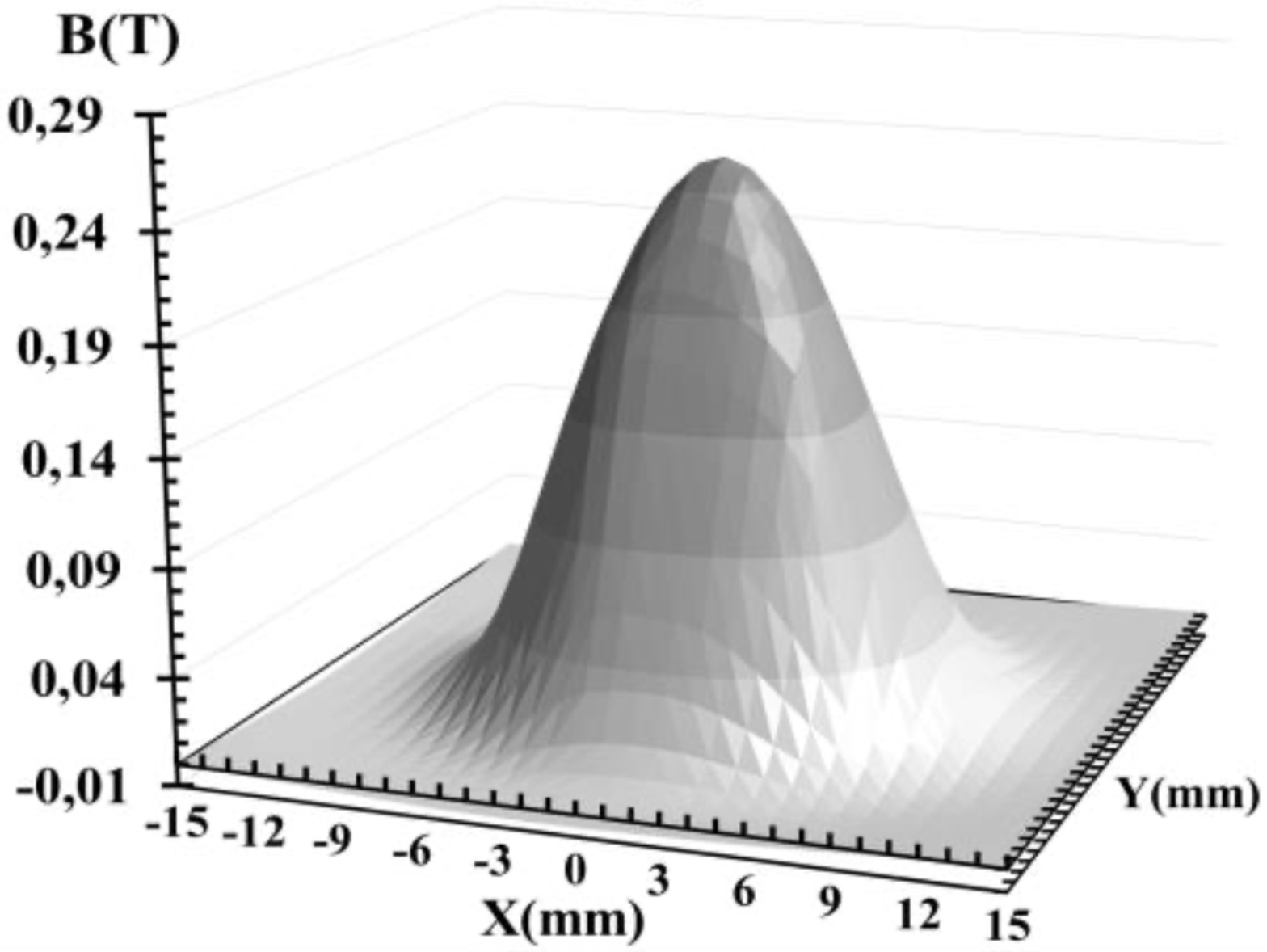}
\caption{\label{fpm2} Magnetic cartography of trapped field ina GdBCO sample magnetised with the flux cooling method at CRISMAT (Caen, France).}
\end{minipage} 
\end{figure}

High temperature superconductors have great advantages for 
energy applications because of their zero electrical resistance 
and relatively low cooling costs. Important devices are
cryo-magnets capable to trap a large magnetic field inside 
a material cooled below a critical temperature. These systems have many applications
such as Maglev trains, motors, wind-mills, etc. Figure \ref{fpm1} illustrates a Pulsed Field magnetization (PFM) setup
where a cooled superconductor is submitted to a short pulse (a few $\mu$s)
of magnetic field.
Figure \ref{fpm2} shows a typical experimental result with a trapped field that is shown to last for days, even weeks \cite{zhou21}.

This multi-physics aspect makes the problem difficult to tackle 
theoretically when flux motion, mechanical and thermal time
scales are considered.
To model PFM, a common pattern is to couple Maxwell's 
equations, a constitutive law, the heat equation and the equations of
elasticity \cite{ainslie}. This type of model permits to reproduce experimental results but does not allow to understand the micro/macroscopic mechanism that enables a superconductor to trap a magnetic field in a few milliseconds.

In this work, we address precisely this question. First, we 
note that Maxwell's equations and the standard constitutive laws
($E/J$ power-law) cannot explain the presence of a stationary field
after the passage of the applied electromagnetic pulse. Then, 
to refine the theoretical approach, we couple Maxwell's equations 
for the vector potential with the Ginzburg-Landau equations. We keep the
wave like character of the model (Lorentz invariance). This is the
Abelian Higgs model.

We study
a  one-dimensional configuration where a superconducting slab is submitted to an
electromagnetic pulse. We derive the equations of motion and interface
conditions using a Lagrangian formalism. The numerical model deals with a pulse longer than the computational domain and absorbing
boundary conditions. Numerical simulations show very little trapping in a
clean sample without defects and important trapping when defects are present.
In the latter case, we observe flux jumps. 

The article is organized as follows: in section 2, we recall the constitutive equations
used for modeling superconductors. In section 3 we introduce the idea of a non trivial fixed point of the system of equations. In section 4, we derive the equations for the  one-dimensional 
system. In section 5 we present the numerical method and in section 6 we show the results.

\section{The model}

Consider Maxwell's equations \cite{feynman}
\begin{eqnarray}
	\label{max1} \nabla \cdot \mathbf{E} = {\rho \over \epsilon_0} ,\\
	\label{max2} \nabla \times \mathbf{E} = -\frac{\partial\mathbf{B}}{\partial t},\\
	\label{max3} \nabla \cdot \mathbf{B}=0,\\
	\label{max4} c^2 \nabla \times \mathbf{B}= {\mathbf{J} \over \epsilon_0} 
	+\frac{\partial\mathbf{E}}{\partial t},
\end{eqnarray}
where $\mathbf{E},\mathbf{B}$ are the electric and magnetic fields, respectively.
$\rho$ is a static charge density, $\mathbf{J}$ is the current (moving charge)
density and $c$ is the velocity of light.

We model a superconducting material. Hence, no
static charges are present and we assume $\rho=0$.
We introduce the vector potential $\mathbf{A}$, such that
\begin{equation}
	\label{vpot} \mathbf{B} = \nabla \times \mathbf{A} .
 \end{equation}
Then Eq. \eqref{max3} is satisfied and Eq. \eqref{max2} implies
\begin{equation}
	\label{vpott} \mathbf{E} = - \frac{\partial\mathbf{A}}{\partial t}.
\end{equation}
We also use the London gauge
\begin{equation}
	\label{lgauge} 
	\nabla \cdot \mathbf{A}=0.
\end{equation}
Then, the last Maxwell equation \eqref{max4} can be written as
\begin{equation}
	\label{At}
\frac{\partial^2{\bf A}}{\partial t^2} -c^2 \Delta \mathbf{A} ={\mathbf{J} \over \epsilon_0} .
  \end{equation}

The main problem in this approach is to find an expression for $\mathbf{J}$.
A first choice is the London approximation \cite{vanduzer}:
\begin{equation}\label{J1}
\mathbf{J} =- {\epsilon_0 c^2 \over \lambda^2} \mathbf{A} ,
 \end{equation}
where $\lambda$ is the London characteristic length.
Another expression, frequently used for large fields is
a power law \cite{rhyner}
\begin{equation}
	\label{J1a}
E =  J^n , 
\end{equation}
where $E$ is the amplitude of $\mathbf{E}$.
This can be refined to obtain the Bean-Kim model \cite{bean}:
\begin{equation}
	\label{J2}
\mathbf{J} =  \mathbf{J_0} {B \over B+B_0}  \tanh({E \over E_c}),
\end{equation}
where $B$ is the magnitude of $\mathbf{B}$ and $J_0, B_0, E_c$
are constants.
\section{Existence of a non trivial fixed point}

Experiments on magnetic flux jumps \cite{weinstein} 
report that there is a threshold of the applied magnetic field above which the
sample switches from a transient to a permanent magnetization.
In \cite{weinstein}, the trapped field is presented as a function of the radius,
for different applied fields $B$ from 1.44 T to 2.69 T, and shows
a clear jump for $B > 1.57$ T.
The trapped field thus produced is observed to remain stable for very long times,
much longer than the duration of the initial pulse.
This indicates that
the system switches from one type of state close to zero to another non-nul state.
Dynamically, this means that there should exists a non trivial 
fixed point in the coupled equations \eqref{At},\eqref{J1} or 
\eqref{At}, \eqref{J2}.

Equations \eqref{At} and \eqref{J1}, or \eqref{At} and \eqref{J2}, do not support non trivial fixed points. This is straightforward to obtain for \eqref{J1}.
For \eqref{J2}, we rewrite the system \eqref{At},\eqref{J2} using \eqref{vpot} and \eqref{lgauge}
\begin{eqnarray}
\displaystyle \frac{\partial\mathbf{A}}{\partial t}= \mathbf{C} , \\
\displaystyle \frac{\partial\mathbf{C}}{\partial t}= c^2 \Delta \mathbf{A} 
	-{1 \over \epsilon_0} \mathbf{J_0} {B \over B+B_0}  \tanh({\parallel \frac{\partial\mathbf{A}}{\partial t}\parallel \over E_c}) ,
\end{eqnarray}
where $\mathbf{C}$ is an auxiliary vector.
Assume all quantities are scalars.
When searching a fixed point, we assume $\displaystyle \frac{\partial{A}}{\partial t}=\frac{\partial{C}}{\partial t}=0$ and find that
the only existing solution is $A=0$.

The conclusion of this simple analysis is that the constitutive
equations \eqref{J1},\eqref{J1a},\eqref{J2} do not explain 
why a trapped field remains after the passage of the pulse. 
To address this issue, we start from first principles
and use the full Ginzburg-Landau functional to describe the 
superconductor.

\section{The Maxwell-Ginzburg-Landau model}

As assumed above, the electromagnetic component of the problem
is described by the vector potential. The superconductivity
is represented by the complex order parameter $\psi$.
The Ginzburg-Landau free energy in SI units can be written 
as \cite{tinkham}
\begin{equation}
	\label{freegl}
F(\mathbf{A}, \psi) = -\alpha |\psi|^2 + {\beta \over 2} |\psi|^4 +
{1 \over 4 m } |(i \hbar \nabla + {2 e} \mathbf{A} )\psi |^2,
\end{equation}
where $m$ is the electron mass and $e$ is the electron charge.
The $\alpha$ coefficient is usually taken as $\alpha_0 (T-T_c)$ where
$\alpha_0 >0$; here we assume $T < T_c$ so we use $-\alpha$.

Most studies of time dependent superconductivity assume that the
system relaxes to an equilibrium. The dynamics is typically of gradient
type. The effect we are describing takes the system out of equilibrium, hence
we need a relativistic extension of the theory of superconductivity.
This is provided by the Abelian Higgs model \cite{abelian}.

For simplicity, we start with a one-dimensional formulation of the problem.
In the spirit of the study \cite{ckm06},
we write a Lagrangian for Maxwell's equations for the single
component $A$ of the vector potential and the Ginzburg-Landau potential 
$${\cal L}(A,\frac{\partial A}{\partial t},\frac{\partial A}{\partial x},\psi,\frac{\partial \psi}{\partial t},\frac{\partial \psi}{\partial x}) = \frac{1}{2 \mu_0 c^2}\left( \frac{\partial A}{\partial t} \right)^2 -  \frac{1}{2 \mu_0}\left( \frac{\partial A}{\partial x}\right)^2
+I(x) \left[ \alpha |\psi|^2 
- {\beta \over 2} |\psi|^4  \right.$$
\begin{equation}
	\label{lag0} \left. + \frac{\hbar^2}{4mc^2}\left|\frac{\partial \psi}{\partial t}\right|^2 - {1 \over 4m} |(i\hbar \nabla +2 e A)\psi|^2 \right]
,\end{equation}
where $I(x)$ is the indicator function of the superconductor.
We introduced a time dependence of $\psi$ following
the Abelian Higgs model \cite{abelian}. 
We use the subscripts $_t,_x$ to indicate partial derivatives.
The $|\frac{\partial \psi}{\partial t}|^2$ term corresponds to a relativistic generalization
of the Ginzburg-Landau theory of superconductors. It
describes a fast and out of equilibrium rearrangement of the 
order parameter.

We introduce the characteristic lengths $\xi, ~\lambda$ and their ratio
$\kappa$ 
\begin{equation}\label{constants}
\xi^2 = { \hbar^2 \over 4 m \alpha},~~
~~\lambda^2 = {m \beta \over 2 e^2 \mu_0 \alpha}, ~~
\kappa = {\lambda \over \xi}.
\end{equation}
Following \cite{as10} we normalize the main variables as
\begin{equation}
	\label{normal}
x= \lambda x',~~t= {\lambda \over c}t', ~~
\psi = \sqrt{\alpha \over \beta}\psi',~~
A= {  \hbar \over 2 e \xi }A' .
\end{equation}

Plugging these expressions into the Lagrangian ${\cal L}$, we obtain
$${\cal L} = \left( \frac{\partial A}{\partial t}\right)^2-\left(\frac{\partial A}{\partial x}\right)^2 
+I(x) \left[ 
{1 \over \kappa^2 }\left|\frac{\partial \psi}{\partial t}\right|^2
- {1 \over \kappa^2 } \left|\frac{\partial \psi}{\partial x}\right|^2
+ i {1 \over  \kappa }  A \left(\psi\frac{\partial \psi^*}{\partial x}-\psi^* \frac{\partial \psi}{\partial x}\right) \right. $$
\begin{equation}
	\label{lagn} 
\left. +|\psi|^2 (1-A^2) - {1 \over 2} |\psi|^4 \right]. 
\end{equation}

The Euler-Lagrange equations 
yield the final system
including the coupling conditions at interfaces $x=\pm L$  (see details in \ref{appA})

\begin{align}
	\label{att1}
	& \frac{\partial^2 A}{\partial t^2} - \frac{\partial^2 A}{\partial x^2}= I(x) \left[  i {1 \over 2 \kappa }   \left(\psi\frac{\partial \psi^*}{\partial x}-\psi^* \frac{\partial \psi}{\partial x}\right)
	-  A |\psi|^2
	\right], \\
	\label{psit1} 
	&  \frac{\partial^2 \psi}{\partial t^2} -  \frac{\partial^2 \psi}{\partial x^2}  = 
	-{i \kappa} \left(  \frac{\partial A}{\partial x} \psi + 2 A \frac{\partial \psi}{\partial x}\right)  
	+ \kappa^2 \psi ( 1 - |\psi|^2 - A^2) = 0,~~~  x \in [-L,L] ,\\
	\label{bound1}
	& - i A \psi + {1 \over \kappa} \frac{\partial \psi}{\partial x}=0 , ~~~x= \pm L  .
\end{align}
Note that the scale of variation of $\psi$ is $1/\kappa$.
For large values of $\kappa$ giant vortex states are
expected \cite{giant1}.

\section{Numerical model}

Partial differential equations \eqref{att1},\eqref{psit1} were
solved using an ODE solver for the time advancement and finite
differences for the space discretization.

\begin{figure}[h]
	\centering
		\includegraphics[height=4.9 cm,width=10.1 cm]{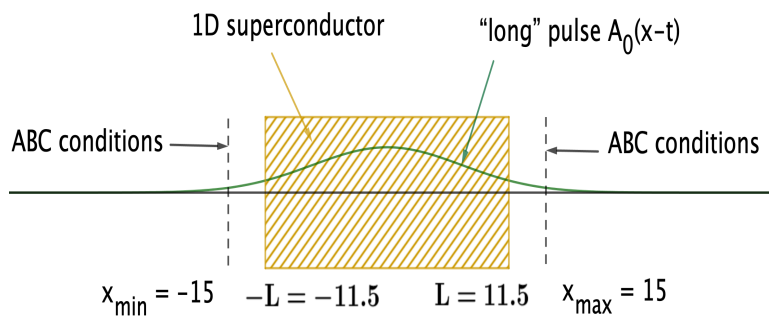}
		\caption{\label{num1D}Schematics of the computational domain.}
\end{figure}

We study how an electromagnetic pulse scatters off a superconducting layer.
It is therefore important to prevent any out-going wave to bounce off
the edge of the computational domain and interact again with the layer.
To prevent this effect, we use absorbing boundary conditions $\displaystyle \frac{\partial A}{\partial t} = - \frac{\partial A}{\partial x}, ~~\frac{\partial A}{\partial t} =  \frac{\partial A}{\partial x}$ 
at the left and right sides of the computational domain, respectively (see Fig. \ref{num1D}).

\subsection{Defects}

To trap magnetic flux inside the superconductor, we also study the effect of including defects
in the model. We speculate that in the superconductor without defects, as soon as the field recedes, the order parameter
returns to its original value and cannot sustain any long term 
magnetization.

Defects can be geometrically modelled  as a wedge placed at the surface of
the superconducting sample \cite{as10}. This is adapted to a
2D modeling of the sample and such a geometrical defect favors the
penetration of vortices inside the sample. Another type of defect
is a material inhomogeneity where the superconductivity breaks down at 
specific locations inside the slab. In practice, this can be obtained 
by bombarding the sample with heavy ions \cite{weinstein}. Our one-dimensional model
then incorporates a function $s(x)$ in the $\alpha$ term of the
free energy \eqref{freegl}.
This term varies from 1 (superconducting)
to -1 (non superconducting), as in \cite{spo17}.

The precise form of the defect is
\begin{equation}\label{defect}
s(x) = \sum_{i=1}^{n_d} H(x-i x_d),
\end{equation}
where $H$ is the square function
$ H(\xi) =-1$ if $\xi \in [-{w_d \over 2},{w_d \over 2}]$, $H(\xi) =1$ otherwise.

The equations including this type of defect are

\begin{align}
	\label{att2}
	& \frac{\partial^2 A}{\partial t^2} - \frac{\partial^2 A}{\partial x^2}= I(x) \left[  i {1 \over 2 \kappa }   \left( \psi\frac{\partial \psi^*}{\partial x}-\psi^* \frac{\partial \psi}{\partial x}\right) 
	-  A |\psi|^2
	\right], \\
	\label{psit2}
	&  \frac{\partial^2 \psi}{\partial t^2} -  \frac{\partial^2 \psi}{\partial x^2}  =
	-{i \kappa} \left( \frac{\partial A}{\partial x} \psi + 2 A \frac{\partial \psi}{\partial x}\right) 
	+ \kappa^2 \psi  \left ( s(x) - |\psi|^2 - A^2 \right ) 
	,\\
	\label{bound2}
	& - i A \psi + {1 \over \kappa} \frac{\partial \psi}{\partial x}=0 , ~~~x= \pm L  .
\end{align}

To model a pulse $A_0$ longer than the computational domain, we use the following transformation:
\begin{equation}
	\label{LongPulse}
	A = A' + A_0(x-t),
\end{equation}
where $\displaystyle A_0(x) = a_0\exp(-\frac{x^2}{2w_0})$.
Since $A_0(x-t)$ satisfies the wave equation with speed $c = 1$, Eqs. \eqref{att2} and \eqref{psit2} become

\begin{align}
	\label{att3}
	&  \frac{\partial^2 A'}{\partial t^2}-  \frac{\partial^2 A'}{\partial x^2}= I(x) \left[  i {1 \over 2 \kappa }   \left( \psi\frac{\partial \psi^*}{\partial x}-\psi^* \frac{\partial \psi}{\partial x}\right) 
	-  (A'+A_0) |\psi|^2
	\right], \\
	\label{psit3}
	&  \frac{\partial^2 \psi}{\partial t^2} -  \frac{\partial^2 \psi}{\partial x^2}  =
	-{i \kappa}\left( \frac{\partial A'}{\partial x} \psi + 2 (A'+A_0) \frac{\partial \psi}{\partial x} \right)
	+ \kappa^2 \psi  \left ( s(x) - |\psi|^2 - (A'+A_0)^2 \right ) 
	,\\
	\label{bound3}
	& - i (A'+A_0) \psi + {1 \over \kappa} \frac{\partial \psi}{\partial x}=0 , ~~~x= \pm L  .
\end{align}

\section{Numerical results}

In all the runs presented in this section, the computational domain is $[-15,15]$
and the superconductor extent is $[-11.5,11.5]$.
The time step is $dt = 10^{-3}$ and the space step $dx= 7.69 ~10^{-3}$.
Most runs were performed with pulses hitting the slab from both
directions. The typical pulse position and width are $x_0=-200$ and  $w=50$, respectively.

\subsection{Modulus and phase }

A preliminary analysis of equations could be useful 
to understand numerical results.
To this purpose we write the equations using modulus and phase of $\psi$:
\begin{equation}\label{psirt}
\psi = \rho e^{i \theta} .
 \end{equation}
Then Eqs. \eqref{att1}-\eqref{psit1} become
\begin{align} \label{att4}
	& \frac{\partial^2 A}{\partial t^2} - \frac{\partial^2 A}{\partial x^2}= I(x)  \rho^2 \left( \frac{1}{\kappa}\frac{\partial \theta}{\partial x} -A\right)   , \\
	\label{rott}
	&  \frac{\partial^2 \rho}{\partial t^2}-  \frac{\partial^2 \rho}{\partial x^2}  = \rho \left(   \left( \frac{\partial \theta}{\partial t}\right) ^2-\left(  \frac{\partial \theta}{\partial x}\right) ^2 \right) 
	+ {2 \kappa} A \rho  \frac{\partial \theta}{\partial x} +  \kappa^2 \rho (1-\rho^2 -A^2) , \\
	\label{tetatt}
	&  \frac{\partial^2 \theta}{\partial t^2}-  \frac{\partial^2 \theta}{\partial x^2}  = {2 \over \rho} \left(   \frac{\partial \theta}{\partial x}  \frac{\partial \rho}{\partial x} - \frac{\partial \theta}{\partial t}  \frac{\partial \rho}{\partial t}\right) 
	- \kappa \frac{\partial A}{\partial x} -2 \kappa A \frac{1}{\rho} \frac{\partial \rho}{\partial x} .
\end{align}
The interface condition at $x=\pm L$ is
\begin{equation}\label{intrtet}
 \frac{\partial \rho}{\partial x} =0, ~~~ \frac{\partial \theta}{\partial x} = \kappa A . 
\end{equation}

When analysing the system \eqref{att4}, \eqref{rott}, \eqref{tetatt} we notice that a first approximation of the trapped field is $ \frac{\partial \theta}{\partial x} =\kappa A \neq 0$.
The corresponding full solution is
\begin{equation}\label{trap}
A = A_\infty , ~~\rho = 1,~~ \frac{\partial \theta}{\partial x} =\kappa A_\infty .
\end{equation}

\subsection{Effect of amplitude $a_0$}

\begin{figure}[h]
	\centering
	\includegraphics[height=4.9 cm,width= 10 cm]{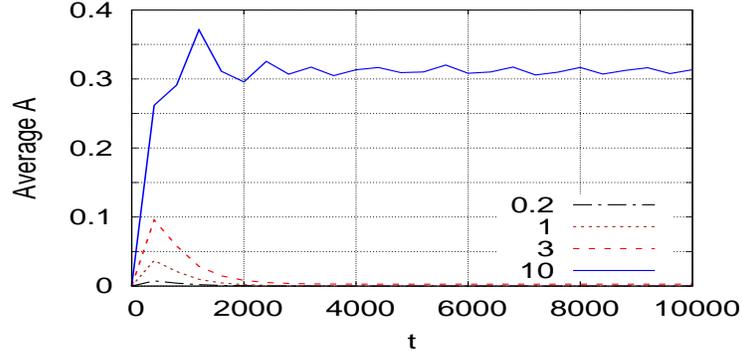}
	\caption{\label{NODEFatrap}Time evolution of the averaged $A$ inside
		the superconducting strip for four amplitudes of the applied pulse: $a_0=0.2,~ 1,~3$ and 10.}
\end{figure}

Figure \ref{NODEFatrap} shows the time evolution of the averaged $A$ inside
the superconducting strip for $a_0=0.2,~ 1,~3$ and 10.
One notes that the trapped $A$ is very small for large amplitudes, $a_0=3$.

Therefore to trap $A$ we need defects that will lock the phase.

\subsection{Influence of defects}

\begin{figure}[h]
	\includegraphics[height=14cm,width=5cm,angle=90]{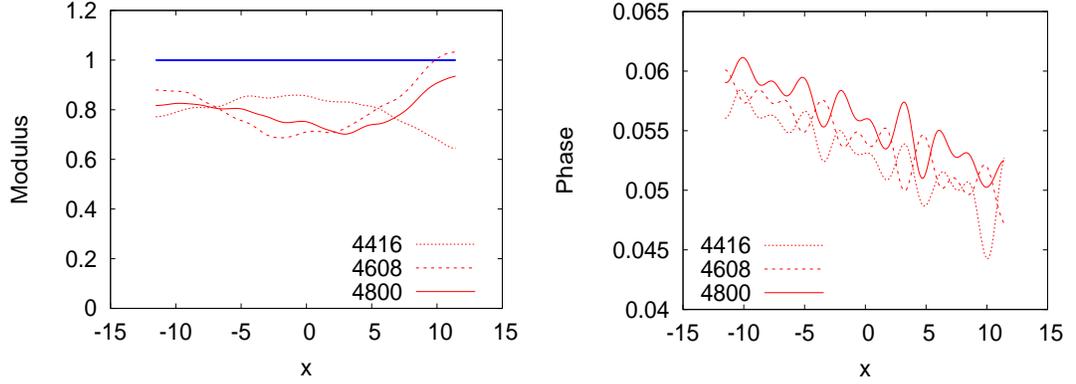}
	\caption{\label{a02}Snapshots of the modulus $\rho(x,t)$ (left panel) 
		and the phase $\theta(x,t)$ (right panel) for $t= 4416, 4608$ and $4800$. The incident pulse
		has amplitude $a_0=0.2$.}
\end{figure}

Figure \ref{a02} shows snapshots of the modulus and phase of $\psi(x,t)$
for $t=4416, 4608$ and $4800$ for an incident pulse of 
amplitude  $a_0=0.2$, with 
defects spaced by 0.05 and without defects. The left panel shows
the modulus of the order parameter $\rho(x,t)$ for both cases. 

Note that without defects $\rho$ equals one (continuous line, blue online) and 
does not evolve. Similarly, the phase $\theta$ remains flat at 0.

\begin{figure}[h]
	\includegraphics[height=4.9 cm,width= 7 cm]{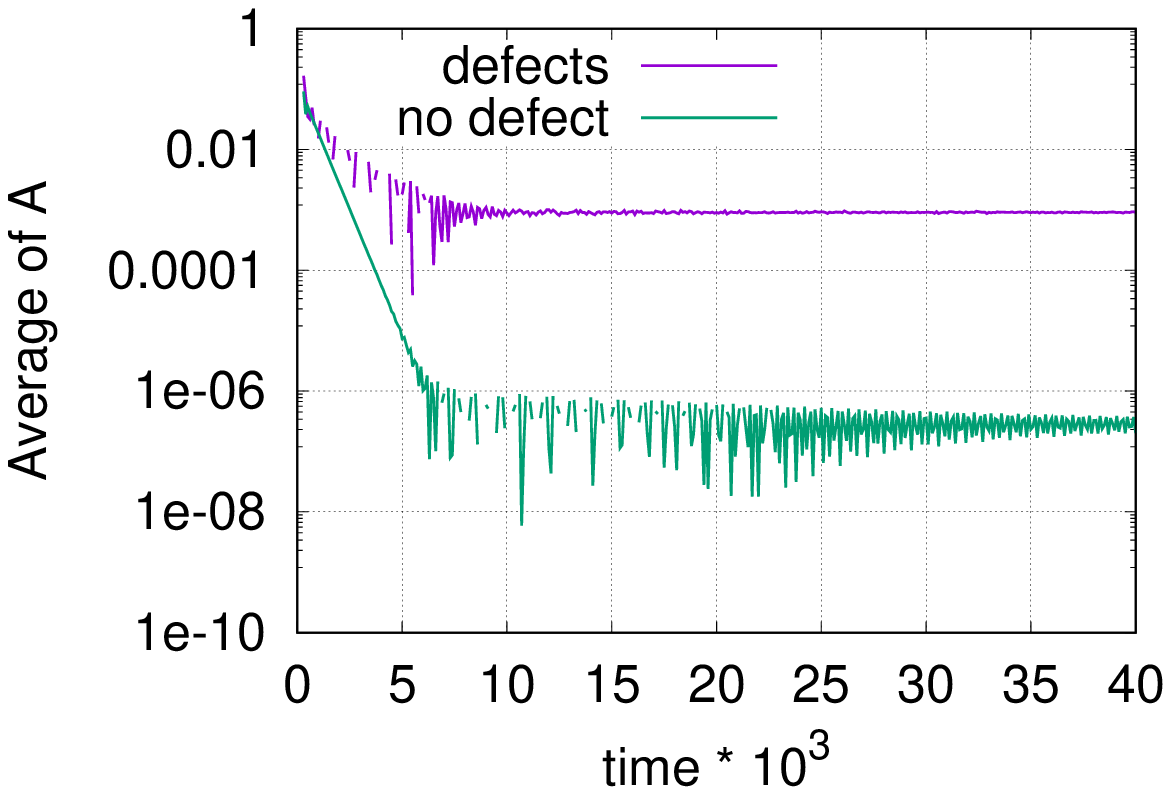}\hfill
		\includegraphics[height=4.9 cm,width= 7 cm]{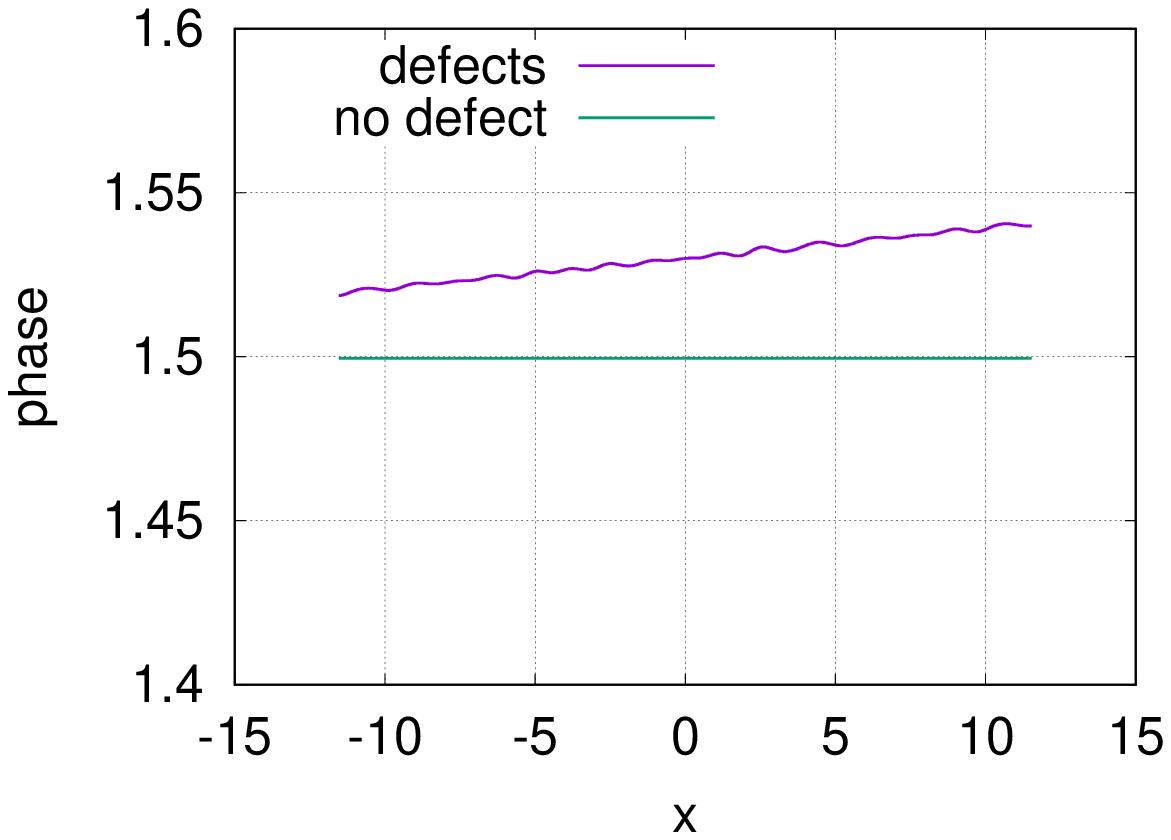}
	\caption{\label{DefectNoDefect}Time evolution of the averaged $A$ (left) and the phase $\theta$ (right) considering the presence or the absence of defects.}
\end{figure}

Figure \ref{DefectNoDefect} shows the importance of defects to obtain a non-zero vector potential.

\subsection {Influence of defect spacing}

\begin{figure}[h]
	\centering
	\includegraphics[height=4.9 cm,width= 10 cm]{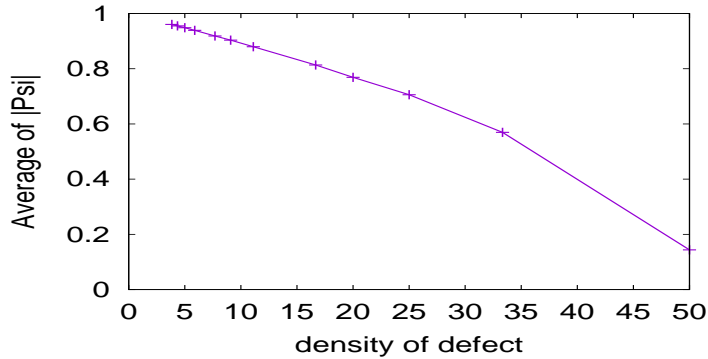}
	\caption{\label{moyMoyPsi}Space and time average of $|\psi|$ as a function of the density of defects (\%).}
\end{figure}

Figure \ref{moyMoyPsi} shows the influence of the density of defects on the modulus of the order parameter. More precisely, the average of $|\psi|$ both on space range $[-L, L]$ and time range $[0, 40000]$ is a decreasing function of the density of defects. We defined this density as the ratio between the number of nodes where $s(x) = -1$ and the ones where $s(x) = +1$.

\begin{figure}[h]
	\includegraphics[height=14cm,width=5cm,angle=90]{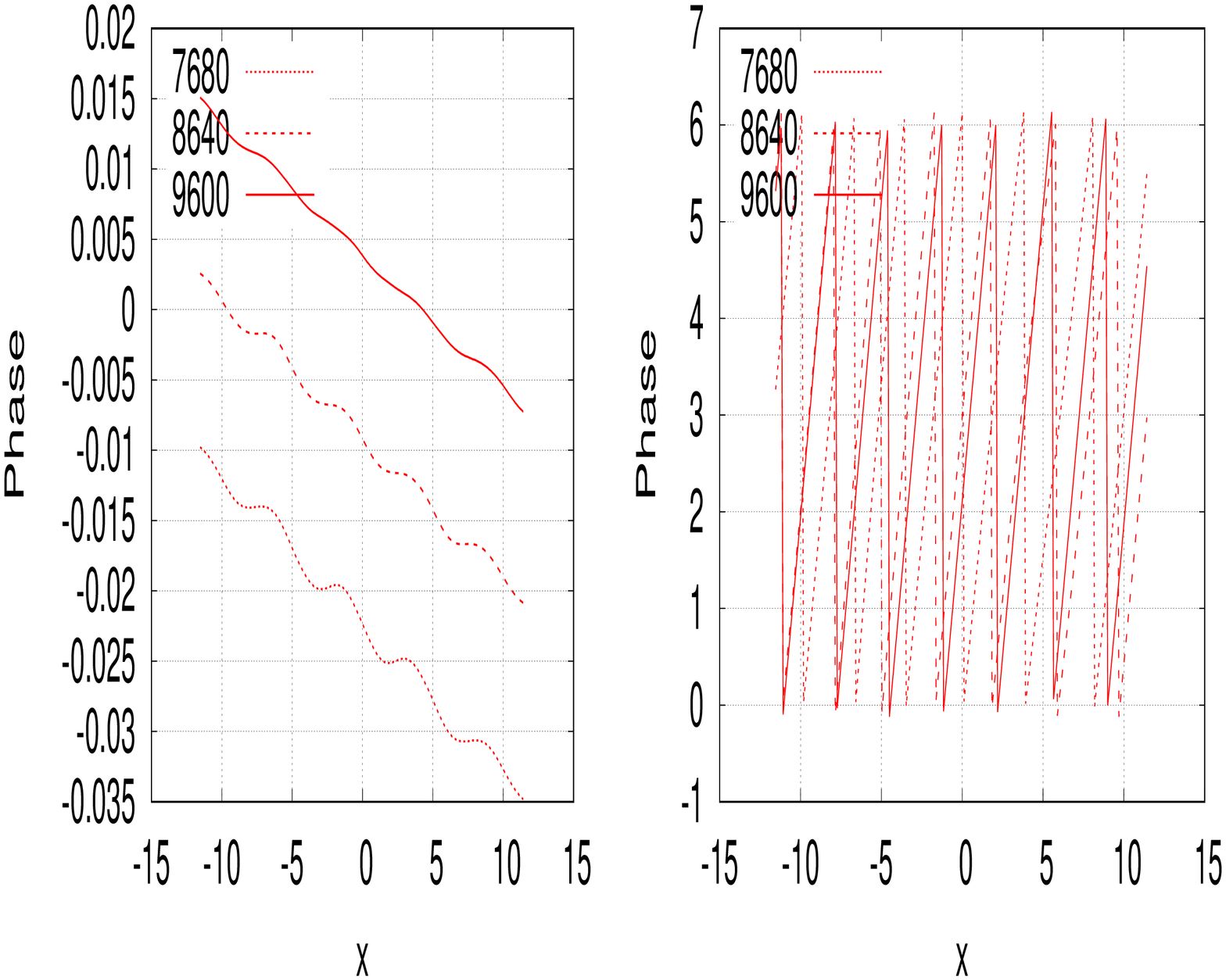}
	\includegraphics[height=14cm,width=5cm,angle=90]{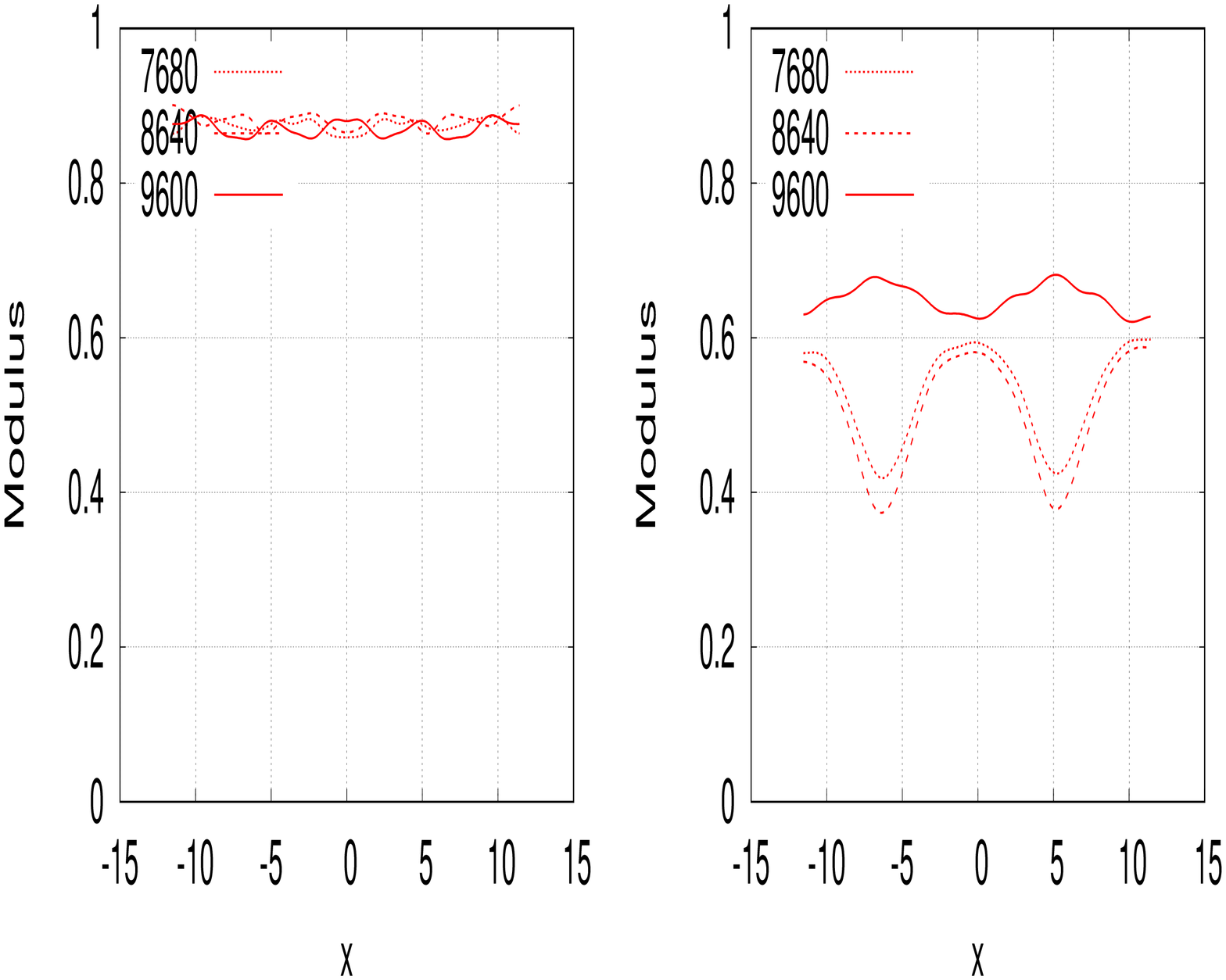}
	\includegraphics[height=14cm,width=5cm,angle=90]{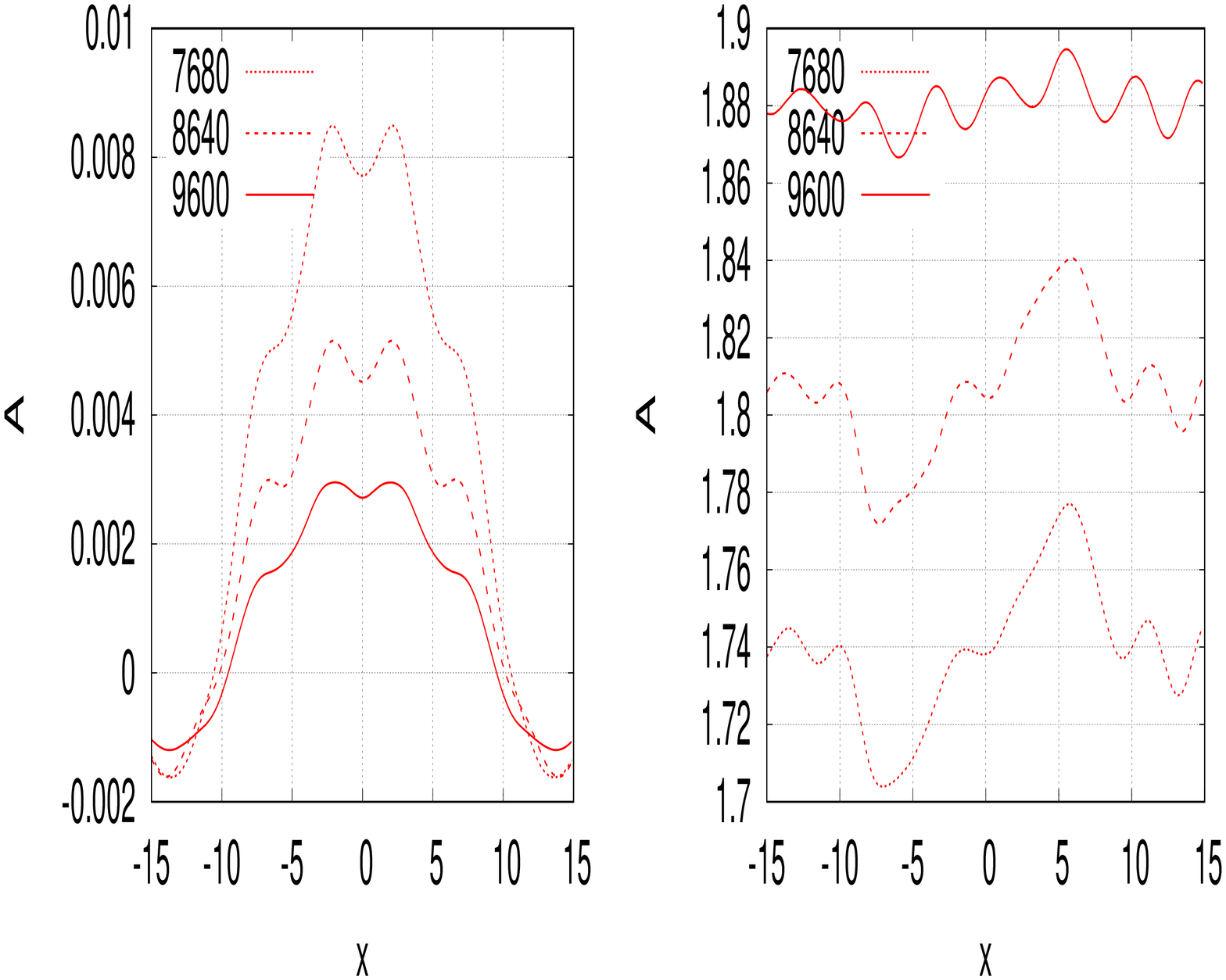}
	\caption{\label{a3ph} Snapshots of the phase $\theta(x,t)$ (top),
		modulus $\rho(x,t)$ (middle) and $A(x,t)$ (bottom) for $t= 7680, 8640$ and 9600
		for a defect spacing $x_d=0.05$ (left panels) and $0.01$ 
		(right panels). The 
		incident pulse has an amplitude $a_0=3$ and the defect spacing is $w_d=10^{-3}$.}
\end{figure}

In Fig. \ref{a3ph} the phase gradient and trapped $A$ are approximately $3\cdot10^{-3}$ for $x_d=0.05$, while they raise to values around $1.8$ for $x_d=0.01$. Moreover, $\displaystyle \frac{\partial \theta}{\partial x}$ is negative for $x_d = 0.05$ and positive for $x_d=0.01$. In addition $\rho(x,t)$ for $x_d=0.01$ becomes rather small and seems to oscillate periodically.

\section{Conclusion}
We presented a one-dimensional model describing the interaction of an electromagnetic pulse with a superconducting slab; it is based on the Abelian Higgs model. The novelty of this approach is to introduce second time derivatives in the equations for both vector potential and order parameter.
The main features of our numerical method are the use of absorbing boundary conditions and an applied pulse whose support is longer than the computational domain.

The equations of motion and boundary conditions are derived from a Lagrangian via Euler-Lagrange equations. Numerical results indicate that the absence of defect leads to a vector potential close to zero. Moreover, the space and time average of the modulus of the order parameter is a decreasing function of the density of defects: this is consistent with the fact that defects reduce superconductivity by pinning vortices and, as a consequence, lower the modulus of the order parameter. 

In conclusion, using this one-dimensional model, we retrieve the usual behaviour of superconductors. Future work will consider 2D configurations.
\clearpage

{\bf Acknowledgements} \\
The present work was performed using computing resources of 
CRIANN (Normandy, France).
The authors thank Nikos Flytzanis, Michikazu Kobayashi  and Mads Peter Soerensen
for useful comments. They also thank Pierre Bernstein and Jacques Noudem 
for sharing their experimental results.

\appendix

\section{Evolution equations and boundary conditions}\label{appA}

The Euler-Lagrange equations are
\begin{eqnarray*}
	\frac{\partial}{\partial t} \left({\partial {\cal L} \over \partial \frac{\partial A}{\partial t} }\right) + 
	\frac{\partial}{\partial x} \left({\partial {\cal L} \over \partial \frac{\partial A}{\partial x} }\right) 
	-{\partial {\cal L} \over \partial A } =0, \\
	\frac{\partial}{\partial t} \left({\partial {\cal L} \over \partial \frac{\partial \psi^*}{\partial t} }\right) + 
	\frac{\partial}{\partial x} \left({\partial {\cal L} \over \partial \frac{\partial \psi^*}{\partial x}}\right) 
	-{\partial {\cal L} \over \partial \psi^* } =0 . 
\end{eqnarray*}
The equation for $A$ is

\begin{equation}  
	\frac{\partial^2 A}{\partial t^2} - \frac{\partial^2 A}{\partial x^2}= I(x) \left[  i {1 \over 2\kappa }   \left( \psi\frac{\partial \psi^*}{\partial x}-\psi^* \frac{\partial \psi}{\partial x}\right)   
	-  A |\psi|^2 \right] .
	\label{ela1}
\end{equation}

For $\psi$ we obtain
\begin{equation}
	\begin{array}{ll}
\displaystyle I(x) \left[ - { 1 \over \kappa^2 }   \frac{\partial^2 \psi}{\partial t^2} + {1 \over \kappa^2}  \frac{\partial^2 \psi}{\partial x^2} 
-{i \over \kappa} \left( \frac{\partial A}{\partial x} \psi + 2 A \frac{\partial \psi}{\partial x}\right) 
+ \psi ( 1 - |\psi|^2 - A^2) \right] \\
\displaystyle	+ \delta(x) \left( -{i \over \kappa} A \psi +{1 \over \kappa^2} \frac{\partial \psi}{\partial x}\right) =0 ,
	\end{array}
		\label{elpsi1} 
\end{equation}

where $\delta(x)$ is the Dirac distribution corresponding to the
$x$ derivative of the characteristic function $I(x)$.

To obtain boundary conditions we integrate \eqref{ela1} over one of the edges of the domain. For example, for $x = -L$ we obtain
$$\int_{-L-\epsilon}^{-L+\epsilon} \frac{\partial^2 A}{\partial t^2} 
-\left[ {\frac{\partial A}{\partial x}  }\right]_{-L-\epsilon}^{-L+\epsilon}
= \int_{-L}^{-L+\epsilon} \left[ i {1 \over 2 \kappa }   \left( \psi\frac{\partial \psi^*}{\partial x}-\psi^* \psi_x\right) 
-  A |\psi|^2 \right] .$$
Taking the limit $\epsilon \to 0$ and assuming bounded variations
of the integrands, we obtain that $\frac{\partial A}{\partial x}$ is continuous at the interface
$x=-L$.
The second equation gives
$$\int_{-L}^{-L+\epsilon} \left[ -{ 1 \over \kappa^2 }   \frac{\partial^2 \psi}{\partial t^2} + {1 \over \kappa^2}  \frac{\partial^2 \psi}{\partial x^2}
+ \dots\right] 
+ \left( -{i \over \kappa} A \psi +{1 \over \kappa^2} \frac{\partial \psi}{\partial x}\right)_{x=-L}=0.$$
Taking the limit $\epsilon \to 0$ and assuming the bracket in 
the integral is bounded, we recover the following standard
boundary condition  \cite{vanduzer} at the edge of a superconductor:

\begin{equation} 
	\label{ju}
	- i A \psi + {1 \over \kappa} \frac{\partial \psi}{\partial x} =0 .
 \end{equation}
\setcounter{section}{1}

\end{document}